\documentclass[aps,pra,reprint,superscriptaddress]{revtex4-1}
\usepackage{blindtext}
\usepackage{amssymb}
\usepackage{amsmath}
\usepackage{tikz}
\usepackage{graphicx}
\usepackage{siunitx}

\newtheorem{prop}{Proposition}\def\PRO{\begin{prop}}\def\ORP{\end{prop}}
\newtheorem{coro}{Corollary}\def\COR{\begin{coro}}\def\ROC{\end{coro}}
\newtheorem{theo}{Theorem}\def\TH{\begin{theo}}\def\HT{\end{theo}}
\def\TH{\begin{theo}}\def\HT{\end{theo}}
\newtheorem{defi}[prop]{Definition}\def\DE{\begin{defi}}\def\ED{\end{defi}}

\newtheorem{lemme}[prop]{Lemma}\def\LE{\begin{lemme}}\def\EL{\end{lemme}}
\usepackage{color}
\usepackage{graphicx,float}
\usepackage{epsfig}
\usepackage{amsmath,amssymb}
\usepackage{bm}
\usepackage[latin1]{inputenc}
\usepackage{hyphenat}
\usepackage[bookmarks=false]{hyperref}
\usepackage{color}
\usepackage[capitalise]{cleveref}
\usepackage{epstopdf}
\usepackage{braket}
\usepackage{verbatim}
\usepackage{comment}
\usepackage{multirow}
\usepackage{float}
\usepackage{subfigure}

\def\ket#1{\left| #1 \right\rangle}

\newcommand{\beq}{\begin{equation}}
\newcommand{\eeq}{\end{equation}}

\definecolor{pink}{RGB}{255,0,255}

\definecolor{ss_color}{rgb}{0,0,1}

\definecolor{darkorange}{RGB}{255,120,0} 

\definecolor{red}{rgb}{1,0,0}

\begin{document}

\title{Upper security bounds for coherent-one-way quantum key distribution}
\author{Javier Gonz\'alez-Payo}
\affiliation{Escuela de Ingenier\'ia de Telecomunicaci\'on, Department of Signal Theory and Communications, University of Vigo, Vigo E-36310, Spain}
\author{R\'obert Tr\'enyi}
\affiliation{Escuela de Ingenier\'ia de Telecomunicaci\'on, Department of Signal Theory and Communications, University of Vigo, Vigo E-36310, Spain}
\author{Weilong Wang}
\affiliation{Escuela de Ingenier\'ia de Telecomunicaci\'on, Department of Signal Theory and Communications, University of Vigo, Vigo E-36310, Spain}
\affiliation{State Key Laboratory of Mathematical Engineering and
Advanced Computing, Zhengzhou, Henan, 450001, China}
\affiliation{Henan Key Laboratory of Network Cryptography Technology, Zhengzhou, Henan, 450001, China}
\author{Marcos Curty}
\email{mcurty@com.uvigo.es}
\affiliation{Escuela de Ingenier\'ia de Telecomunicaci\'on, Department of Signal Theory and Communications, University of Vigo, Vigo E-36310, Spain}

\date{\today}

\begin{abstract}
The performance of quantum key distribution (QKD) is severely limited by multi-photon pulses emitted by laser sources due to the photon-number splitting attack. Coherent-one-way (COW) QKD has been introduced as a promising solution to overcome this limitation, and thus extend the achievable distance of practical QKD. Indeed, thanks to its experimental simplicity, the COW protocol is already used in commercial applications. Here, we derive simple upper security bounds on its secret key rate, which demonstrate that it scales at most quadratically with the system's transmittance, thus solving a long-standing problem. That is, in contrast to what has been claimed, this approach is inappropriate for long-distance QKD transmission. Remarkably, our findings imply that {\it all} implementations of the COW protocol performed so far are insecure. 
\end{abstract}

\maketitle

\noindent {\it Introduction.}---Quantum key distribution (QKD)~\cite{qkd1,qkd2} allows two distant parties (Alice and Bob) to distribute an information-theoretic secure secret key. Due to channel loss, however, the key rate of point-to-point QKD is fundamentally limited, and scales at most linearly with the system's transmittance $\eta$~\cite{TGW,PLOB}. This limitation could be overcome by using intermediate nodes together with, say, twin-field QKD~\cite{TF1,TF2,TF3,TF4}, satellite to ground links~\cite{sat1,sat2}, or, in the long term, quantum repeaters~\cite{QR1,QR2,QR3}.

Besides channel loss, device imperfections also severely limit the performance of practical QKD. One main imperfection are multi-photon signals emitted by laser sources generating weak coherent pulses (WCPs) due to the photon-number-splitting (PNS) attack~\cite{pns1,pns2}. As a result, the secret key rate of the standard BB84 protocol~\cite{BB84} with WCPs is of order $O(\eta^2)$~\cite{pnsScaling}.  

To enhance the performance of point-to-point QKD, there are three main approaches. The first one is decoy-state QKD~\cite{decoy1,decoy2,decoy3}. The second solution uses strong reference pulses~\cite{strong1,strong2,strong3}. Both options deliver a key rate of order $O(\eta)$, thus matching the best possible scaling. The third approach is distributed-phase-reference (DPR) QKD~\cite{dps1,dps2,dps3,dps5,dps6,dps4,dps7,dps8,cow1,cow2,cow2b,cow3,cow4}. One prominent example here is coherent-one-way (COW) QKD~\cite{cow1,cow2,cow2b,cow3,cow4}, which uses a simple experimental setup and is probably robust against the PNS attack. Indeed, long-distance demonstrations (beyond 300 km) of the COW protocol have been reported recently~\cite{cow4}, and this scheme is even used in commercial setups~\cite{IdQ}. Despite these promising results, however, its actual performance has not been fully established yet. There is a big gap between known lower security bounds, whose key rate is of order $O(\eta^2)$ against general attacks~\cite{low_cow} and of order $O(\eta)$ against a restricted type of collective attacks~\cite{cow4}, and the upper security bounds that predict a key rate of order $O(\eta)$~\cite{upp_cow1}.

In this Letter, we close this gap by providing ultimate upper security bounds for the COW scheme, demonstrating a key rate at most of order $O(\eta^2)$, thus matching the scaling of the lower security bounds in~\cite{low_cow}. That is, in contrast to what has been claimed, this protocol is inappropriate for long-distance QKD transmissions. We use a special type of intercept-resend attacks, the so-called sequential attacks~\cite{seq1,seq2,seq3}, which are particularly suited to attack DPR QKD. They transform the quantum channel into an entanglement breaking channel and, thus, no secret key can be generated~\cite{condition}. In doing so, we explicitly show that {\it all} implementations of the COW scheme performed so far are insecure, which is remarkable.

\begin{figure}
\centering{\includegraphics*[scale=0.44]{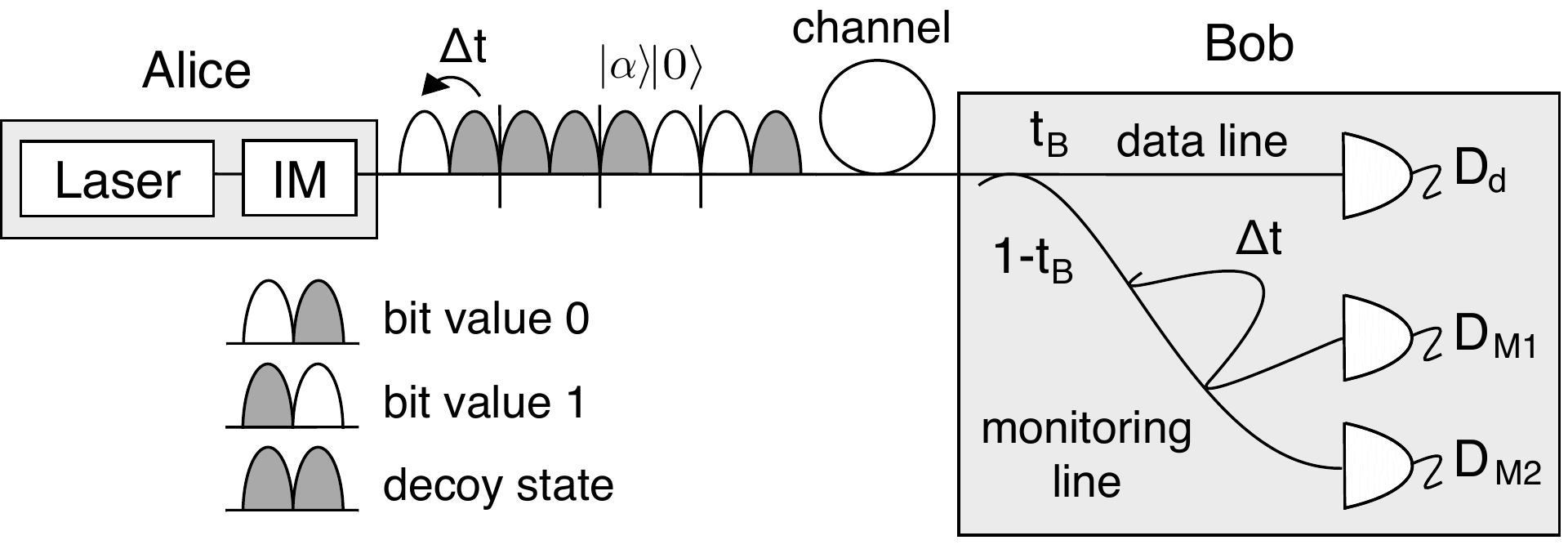}}
\caption{Schematic description of the COW protocol. Alice sends Bob coherent states $\ket{0}\ket{\alpha}$, $\ket{\alpha}\ket{0}$ and $\ket{\alpha}\ket{\alpha}$ that she selects at random each given time. Bob passively distributes the incoming signals into the data and the monitoring lines by means of a beamsplitter of transmittance $t_{\rm B}$. The monitoring line consists of a Mach-Zehnder interferometer that measures the coherence between adjacent pulses. In the figure: IM, intensity modulator; $\Delta{}t$, time delay between adjacent pulses; D$_{\rm d}$, D$_{\rm M1}$ and D$_{\rm M2}$, single-photon detectors.}
\label{COW}
\end{figure}

\noindent {\it Coherent-one-way QKD~\cite{cow1,cow2,cow2b,cow3,cow4}.}---The basic setup is illustrated in Fig.~\ref{COW}. Alice uses a laser source, together with an intensity modulator, to generate a sequence of coherent states $\ket{0}\ket{\alpha}$, $\ket{\alpha}\ket{0}$ and $\ket{\alpha}\ket{\alpha}$ that she sends to Bob, where $\ket{0}$ is the vacuum state. These signals correspond, respectively, to a bit value $0$, a bit value $1$, and a decoy signal. They are generated with {\it a priori} probabilities $P_{0}=P_{1}=(1-f)/2$ and $P_{\rm d}=f$, respectively, for a given $f$. At Bob's side, a beamsplitter of transmittance $t_{\rm B}$ distributes the incoming signals into the data and the monitoring lines. The data line discriminates the bit states $\ket{0}\ket{\alpha}$ and $\ket{\alpha}\ket{0}$ by measuring the arrival time of each signal with the detector D$_{\rm d}$. The monitoring line measures the coherence between adjacent non-empty pulses to check for eavesdropping. This is done with a Mach-Zehnder interferometer followed by two detectors, D$_{\rm M1}$ and D$_{\rm M2}$. The interferometer is arranged such that adjacent coherent states $\ket{\alpha}$ cannot produce a ``click'' in say detector D$_{\rm M2}$. Errors in the data line are characterized by the quantum bit error rate (QBER), while those in the monitoring line are quantified with the visibilities
\begin{equation}
{V_s} = \frac{{p_{\rm click}({{\rm{D_{M1}}}|s}) - p_{\rm click}({{\rm{D_{M2}}}|s})}}{{p_{\rm click}({{\rm{D_{M1}}}|s}) + p_{\rm click}({{\rm{D_{M2}}}|s})}}, \label{vis_gen}
\end{equation}
with $s\in{\mathcal S}=\{``d$''$, ``01$''$, ``0d$''$, ``d1$''$, ``dd$"$\}$. Here, $p_{\rm click}({\rm D}_{{\rm M}i}|s)$ is the conditional probability that detector ${\rm D}_{{\rm M}i}$ ``clicks'' given that the two adjacent states $\ket{\alpha}$ come from a sequence $s$ sent by Alice. A sequence $s=``d$'' corresponds to a decoy signal ({\it i.e.}, $\ket{\alpha}\ket{\alpha}$), a sequence $s=``01$'' corresponds to a bit 1 signal followed by a bit 0 signal ({\it i.e.}, $\ket{0}\ket{\alpha}\ket{\alpha}\ket{0}$), and the other sequences $s\in{\mathcal S}$ are defined similarly. 

Alice and Bob form the sifted key with those instances where Alice emitted a bit state $\ket{0}\ket{\alpha}$ or $\ket{\alpha}\ket{0}$ and Bob obtained a ``click'' in D$_{\rm d}$. Next, they use error correction (based on their estimate of the QBER), and privacy amplification (which depends on the estimated visibilities $V_s$) to distill a final secure key from the sifted key. 

\noindent {\it Upper security bounds.}---So far, upper bounds on the secret key rate of the COW protocol have been established by considering mainly collective attacks~\cite{upp_cow1}. That is, they assume that the eavesdropper, Eve, attacks each of Alice's signals independently from each other, and she uses the same {\it individual} strategy to interact with each of them. This approach provides upper bounds that scale linearly with $\eta$~\cite{upp_cow1}.

Below, we derive significantly tighter upper bounds on the performance of the COW scheme by using sequential attacks~\cite{seq1,seq2,seq3}. Importantly, in a sequential attack Eve decides {\it jointly} the signals that she sends to Bob. That is, such signals can now depend on {\it all} the measurement results obtained by Eve after measuring all the signals emitted by Alice. This property makes these attacks particularly suited to attack DPR QKD. Indeed, sequential attacks have been already successfully applied against differential-phase-shift (DPS) QKD~\cite{dps1,dps2,dps3} in~\cite{seq1,seq2,seq3}. A key limitation of these attacks when used against DPS QKD is, however, that they introduce errors, which restrict their effectiveness. This is so because in DPS QKD any break in the coherence between adjacent pulses can result in an error. In the Supplemental Material, we use a sequential attack to derive explicit upper bounds on the key rate of the DPS protocol that scale as $O(\eta^\gamma)$, where the parameter $\gamma>1$ depends on the observed error rate.

The main observation of this paper is, however, to note that sequential attacks can be very effective against the COW protocol. This is due to the combination of two special properties of Alice's signals in this scheme. First, they are linearly independent. This means that Eve could use an unambiguous state discrimination (USD) strategy~\cite{chefles_usd1,chefles_usd2,eldar1} to distinguish $\ket{0}\ket{\alpha}$, $\ket{\alpha}\ket{0}$ and $\ket{\alpha}\ket{\alpha}$ without introducing any error~\cite{footnote}. And, second, Alice's signals contain the vacuum state, which naturally breaks the coherence between adjacent pulses. That is, in contrast to DPS QKD, here Eve could send Bob blocks of signals, separated by vacuum states, without introducing errors. Indeed, Eve only needs to maintain the coherence between those consecutive non-empty pulses sent by Alice that she resends to Bob. As we will show below, thanks to these two properties together, it turns out that for any value of $\alpha$, there is always a loss regime in which the signals sent by Eve result in QBER=0 and $V_s=1$ for all $s\in{\mathcal S}$ despite the fact that sequential attacks do not allow the distribution of a secure key~\cite{condition}. 
   
Next, we describe briefly a slightly simplified version of the sequential attack that we consider, which already captures its main features. A detailed description of the attack and the parameters that Eve can tune can be found in the Supplemental Material. In particular, Eve first measures each signal emitted by Alice with a measurement strategy which resembles that introduced in~\cite{chefles_mixed,meas_mixed}. That is, it lies between the so-called minimum error discrimination (MED) strategy~\cite{Hel,chefles}, and the USD strategy~\cite{chefles_usd1,chefles_usd2,eldar1}. Each measurement provides Eve four possible outcomes: either it identifies Alice's state or it provides an  inconclusive result. The probability of this latter event, which we shall call $q_{\rm inc}$, depends on the overall system loss, and is selected {\it a priori} by Eve to reproduce the expected gain at Bob's data line ({\it i.e.}, the probability that Bob observes a detection event per signal sent by Alice). For any $q_{\rm inc}$, Eve's measurement minimizes the error probability to distinguish Alice's states conditioned on outputting a conclusive result. That is, when $q_{\rm inc}=0$ ($q_{\rm inc}\geq{}q_{\rm usd}$), her measurement matches the MED (USD) strategy, with $q_{\rm usd}<1$ being the failure probability of the optimal USD measurement able to distinguish Alice's states. See the Supplemental Material for a detailed description of Eve's measurement. 

Once Eve has measured all the signals emitted by Alice, she prepares new signals that she sends to Bob. For a given value of the gain at Bob's side, her goal is to minimize (maximize) the QBER (visibilities $V_s$). For this, she proceeds as follows. Whenever her measurement result is inconclusive, or the number of consecutive conclusive measurement results is below a certain threshold value, say $M_{\rm min}$, Eve sends Bob vacuum signals to avoid errors. Note that in the conservative untrusted device scenario, where Eve can modify the parameters of Bob's detectors (particularly their detection efficiency and dark count rate), vacuum signals do not produce a ``click'' at Bob's side. On the other hand, if the number of consecutive conclusive measurement results is greater than, or equal to, $M_{\rm min}$, she sends Bob a sequence of signals that may contain non-empty pulses via a lossless channel. These signals correspond mainly to the results obtained with her measurements but could be slightly adjusted. Precisely, depending on the expected gain, Eve may optimize the intensity $|\beta|^2$ of the non-empty pulses she sends Bob to enhance the probability that he detects them. Moreover, Eve might slightly process each sequence of signals before she sends it to Bob to increase the resulting visibilities, as we explain below. This is because even if a sequence of signals prepared by Eve perfectly matches that emitted by Alice (except for the intensity), only those sequences whose first (last) signal is $\ket{\beta}\ket{0}$ ($\ket{0}\ket{\beta}$) can guarantee perfect visibility results at Bob's side. As already explained, note that  Alice and Bob only check coherence between adjacent non-empty pulses, and the presence of a vacuum state in $\ket{\beta}\ket{0}$ ($\ket{0}\ket{\beta}$) breaks the coherence between this signal and the preceding (following) one. This means, in particular, that to improve the visibilities Eve should favor the transmission of those blocks of signals which start (end) with the signal $\ket{\beta}\ket{0}$ ($\ket{0}\ket{\beta}$). For this, with probability $q_{\rm p}$ Eve selects the largest subsequence of signals (within the original sequence) whose first (last) signal is $\ket{\beta}\ket{0}$ ($\ket{0}\ket{\beta}$) and sends this sequence to Bob (after adding the necessary vacuum states), while, with probability $1-q_{\rm p}$, she directly sends Bob the original sequence of signals without making any adjustment. In doing so, Eve can use the parameter $q_{\rm p}$ to tune the gain and visibilities observed by Bob. Precisely, by increasing $q_{\rm p}$ Eve can decrease (increase) Bob's gain (visibilities $V_s$). 
\begin{figure}[htbp]
\centering{\includegraphics*[scale=0.43]{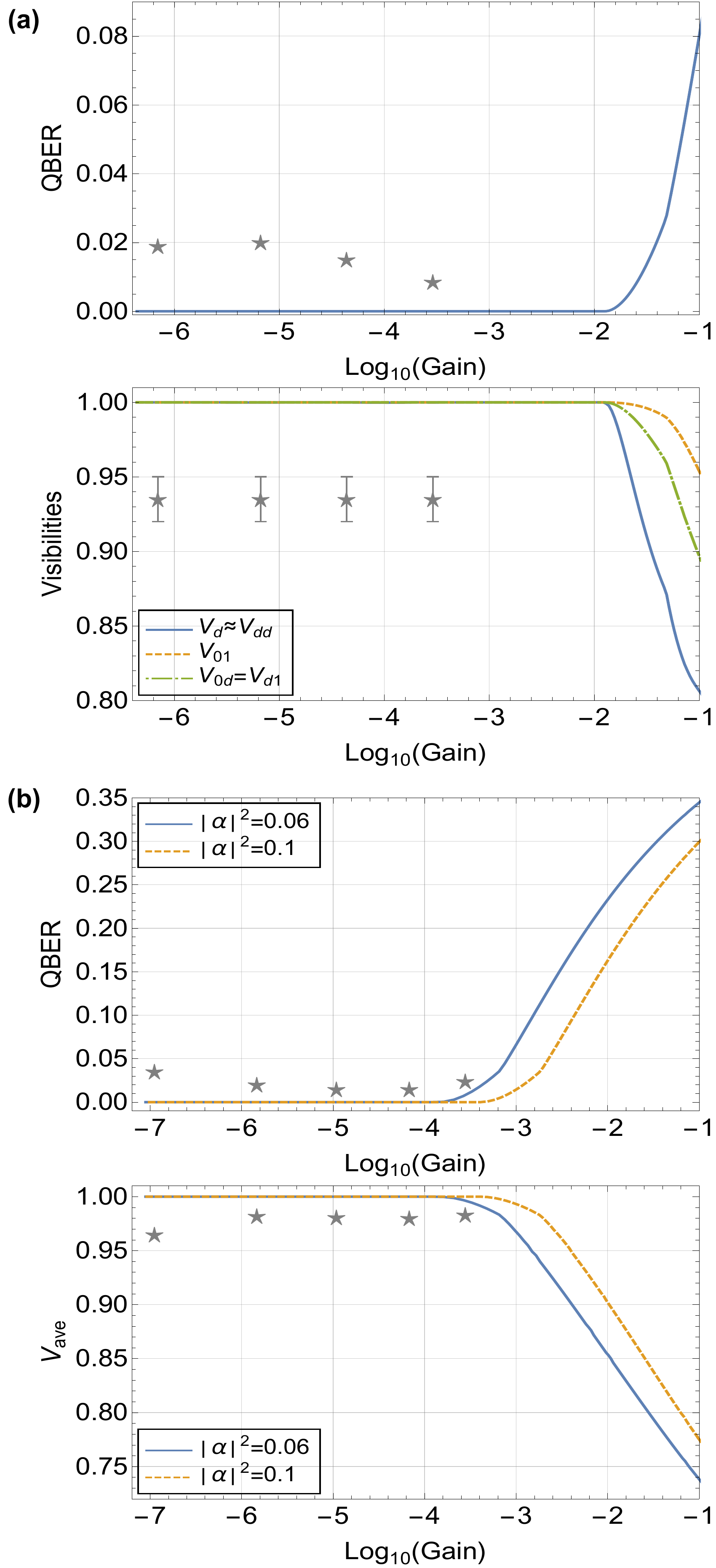}}
\caption{QBER and visibilities versus gain at Bob's side in the sequential attack considered. The stars represent experimental data from~\cite{cow3} (Subfigure (a)) and~\cite{cow4} (Subfigure (b)). In Subfigure (a) the uncertainty bars match the information provided in~\cite{cow3}. In Subfigure (b), $V_{\rm ave}$ refers to the average visibility~\cite{Vave} used in~\cite{cow4}. Also, the two lines correspond to those experiments using the weakest and the strongest intensity $|\alpha|^2$. For other experiments, the results lie exactly between these two lines, and are omitted for simplicity.}
\label{fig:Experiments}
\end{figure}

We find, therefore, that for those values of the gain which can be reproduced by Eve by selecting $q_{\rm inc}\geq{}q_{\rm usd}$ and $q_{\rm p}=1$, her attack achieves QBER=0 and $V_s=1$ for all $s\in{\mathcal S}$. We call this scenario the perfect USD regime. For higher values of the gain, Eve can optimize the parameters of her attack to minimize (maximize) the QBER (visibilities $V_s$) observed by Bob. We refer the reader to the Supplemental Material for analytical expressions of the resulting gain, QBER and visibilities $V_s$ as a function of Eve's parameters. 

\noindent {\it Evaluation.}---We first apply the sequential attack introduced above to the long-distance experimental implementations of the COW protocol reported in~\cite{cow3,cow4}. Afterward, we evaluate a simple upper bound on its secret key rate, $K$, which demonstrates a scaling of order $O(\eta^2)$. 

In Fig.~\ref{fig:Experiments} we show the QBER and visibilities which are achievable by Eve, as a function of the gain at Bob's side, for the experiments in~\cite{cow3,cow4}. For this, we optimize numerically the analytical expressions that describe Eve's attack over all parameters that she controls. Precisely, in Fig.~\ref{fig:Experiments}(a) (Fig.~\ref{fig:Experiments}(b)) we maximize the minimum value of all the visibilities $V_s$ (the average visibility $V_{\rm ave}$). The average visibility $V_{\rm ave}$ is defined by using a weighted combination of the conditional detection probabilities $p_{\rm click}({\rm D}_{{\rm M}i}|s)$~\cite{Vave}, and is the quantity considered in~\cite{cow4}. Due to the symmetry of Alice's signal states as well as Eve's attack, it turns out that $V_{0d}=V_{d1}$. Also, in Fig.~\ref{fig:Experiments}(a) we find that $V_{d}\approx{}V_{dd}$, though this is not true in general. As expected, when the gain at Bob's side decreases, the achievable QBER by Eve also decreases and the visibilities $V_s$ increase, till they reach the perfect USD regime where QBER$=0$ and $V_s=1$ for all $s\in{\mathcal S}$. Importantly, this regime is reached very rapidly by Eve ({\it i.e.}, for relatively large values of the gain) unless Alice selects $\alpha$ very small (which in turn significantly reduces the secret key rate). Indeed, by decreasing $\alpha$, Alice's signals become less orthogonal to each other, and thus Eve must increase $q_{\rm inc}$ to be able to distinguish them unambiguously. In turn, this reduces the maximum value of the gain at which the perfect USD regime is possible. Each experiment in~\cite{cow3,cow4} is indicated in Fig.~\ref{fig:Experiments} with a star symbol. The covered distances range from 100 km to 250 km (104 km to 307 km) in~\cite{cow3} (\cite{cow4}). All experimental parameters can be found in the Supplemental Material. Importantly, Fig.~\ref{fig:Experiments} shows that all the implementations in~\cite{cow3,cow4} are insecure, as their experimental QBER value (visibilities) is (are) above (below) the achievable values by Eve's sequential attack. Remarkably, the same conclusion applies to all other realizations of the COW protocol~\cite{cow2,cow2b} (see the Supplemental Material for further details). That is, our findings imply that no secure implementation of COW has been reported yet.

Similarly, if we compare the performance of the sequential attack above with the upper bounds derived in~\cite{upp_cow1}, which only consider collective two-pulse attacks, it can be shown that~\cite{upp_cow1} significantly overestimates the gain region where the COW protocol could be secure at all. See the Supplemental Material for further details. 
\begin{figure}
\centering{\includegraphics*[scale=0.37]{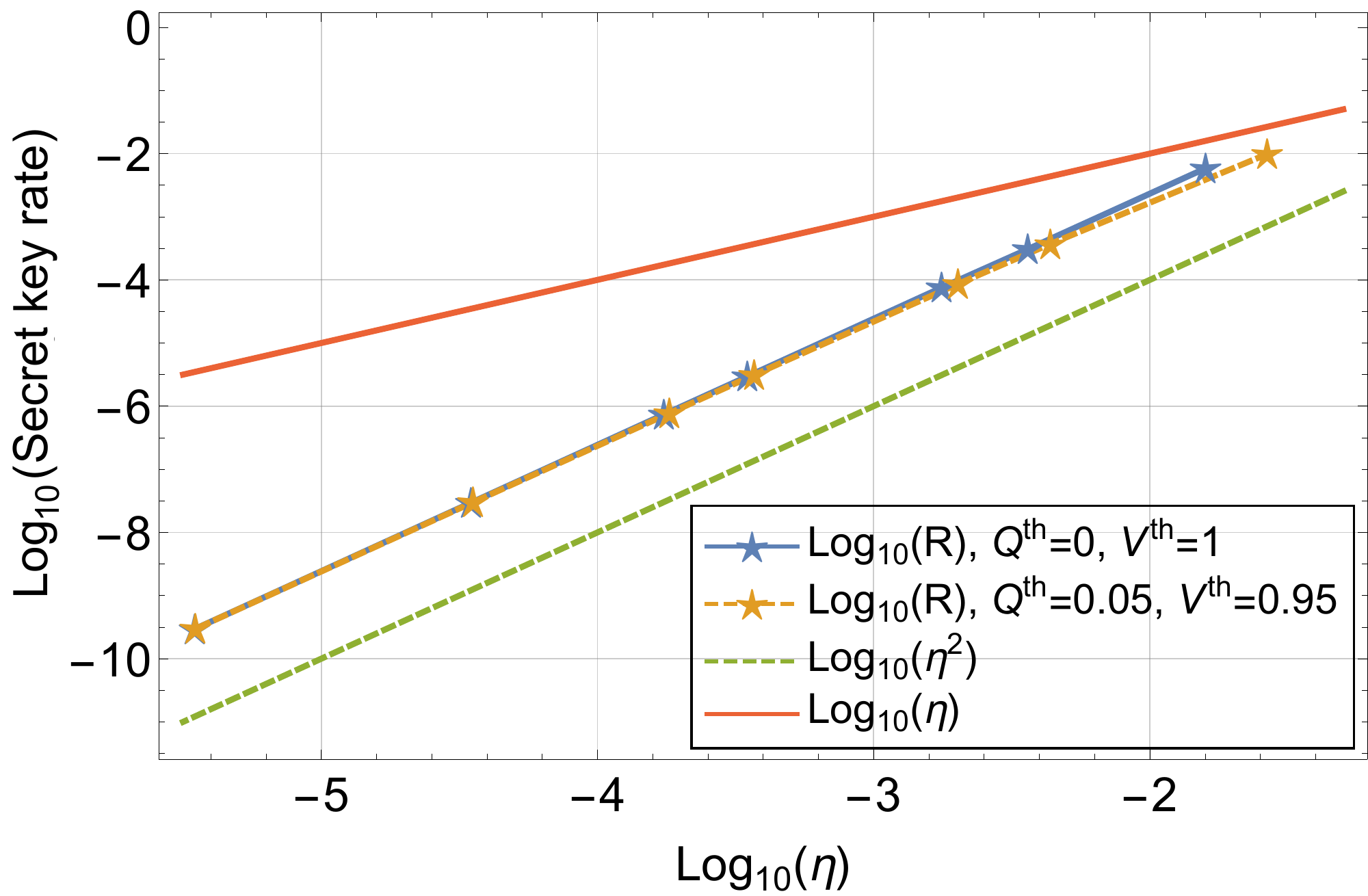}}
\caption{Upper bound $R$ on the secret key rate $K$ of the COW protocol as a function of the system's transmittance $\eta$ when $f=0.155$~\cite{cow4}. The value of $R$ corresponding to the case $Q^{\rm th}=0$ and $V^{\rm th}=1$ ($Q^{\rm th}=0.05$ and $V^{\rm th}=0.95$) is illustrated by blue (yellow) stars. Both cases basically overlap each other for small values of $\eta$. For comparison, the figure includes as well the lines equal to $\eta$ (red solid line) and $\eta^2$ (green dashed line).}
\label{fig:UpperBound}
\end{figure}

Finally, we use the sequential attack above to obtain an explicit upper security bound on the secret key rate $K$ of the COW protocol. Our starting point is a trivial upper bound on $K$, which is the probability that Alice sends Bob a bit signal state and he observes a ``click'' in his data line. That is, it holds that $K\leq{}(1-f)[1-\exp{(-\eta{}t_{\rm B}|\alpha|^2)}]<(1-f)\eta{}|\alpha|^2$, where in the second inequality we use the fact that $[1-\exp{(-\eta{}t_{\rm B}|\alpha|^2)}]\leq{}\eta{}t_{\rm B}|\alpha|^2<\eta{}|\alpha|^2$, since we have that $t_{\rm B}<1$. Next, we determine the maximum possible value of $|\alpha|^2$, which we shall call $|\alpha_{\rm max}(f)|^2$, because it depends on $f$. Precisely, for a given value of the gain at Bob's side and for given threshold values of the QBER and visibilities, say $Q^{\rm th}$ and $V^{\rm th}$, we optimize numerically the maximum intensity $|\alpha_{\rm max}(f)|^2$ such that the sequential attack above is unsuccessful. We say that the sequential attack is unsuccessful if it does not simultaneously satisfy QBER$\leq{}Q^{\rm th}$ and $V_s\geq{}V^{\rm th}$ for all $s\in{\mathcal S}$. Conversely, this implies that if $|\alpha|^2>|\alpha_{\rm max}(f)|^2$ then Eve's attack is successful, which we confirmed with our simulations. Given $|\alpha_{\rm max}(f)|^2$, we have that
\begin{equation}
K<(1-f)\eta{}|\alpha_{\rm max}(f)|^2\equiv{}R.
\end{equation}
This upper bound is illustrated in Fig.~\ref{fig:UpperBound} for the cases $Q^{\rm th}=0$ and $V^{\rm th}=1$, and $Q^{\rm th}=0.05$ and $V^{\rm th}=0.95$, and assuming $f=0.155$~\cite{cow4}. When $f=0.0625$~\cite{cow3}, the resulting upper bounds are marginally lower than those shown in Fig.~\ref{fig:UpperBound}, and almost overlap with each other. The values of $|\alpha_{\rm max}(f)|^2$ can be found in the Supplemental Material; importantly, it can be shown that they decrease linearly with the system's transmittance $\eta$ and have the same slope for any $f\in(0,1)$. We find, therefore, that $R$ scales at most quadratically with $\eta$ (see Fig.~\ref{fig:UpperBound}). 

\noindent {\it Conclusion.}---We have derived simple upper security bounds for coherent-one-way (COW) quantum key distribution (QKD). They exploit sequential attacks and the fact that Alice's signals are linearly independent and, moreover, they contain vacuum states, which naturally break the coherence between adjacent pulses. By using a simple eavesdropping strategy, we have explicitly shown that all implementations of the COW protocol reported so far appear to be insecure. Most importantly, our results demonstrate that the key rate of this protocol scales at most quadratically with the system's transmittance, given that Alice and Bob only monitor the QBER and visibilities. This renders this scheme inappropriate for long-distance quantum communications. 

To enhance the robustness of COW against sequential attacks, one could either monitor more observables ({\it e.g.} the detection rates of Alice's signals), and/or measure the coherence between non-adjacent pulses, following similar ideas like in~\cite{dps4}. The first approach requires the development of novel security proof techniques that take this additional information into account. This might be challenging because Eve could attack only certain blocks of signals, or exploit the inevitable statistical fluctuations in a finite-key regime to slightly deviate from the expected detection rates. The second approach needs a much more cumbersome receiver, thus eliminating the main advantage of COW, {\it i.e.}, its simple experimental setup. 

\noindent {\it Note added.}---We shared the finished manuscript with ID Quantique before its submission for publication. The company stated that their commercial QKD products based on the COW protocol run on relatively short distances and use a sufficiently weak mean number of photons, such that they can detect sequential attacks. However, the value of mean number of photons employed would not allow ID Quantique products to detect sequential attacks if the distances were much larger than their present limit.

\noindent {\it Acknowledgments.}--- The authors wish to thank C. C. W. Lim, H.-K. Lo, B. Korzh, C. Pacher, N. Walenta, and H. Zbinden for very useful discussions. This work was funded by the Galician Regional Government (consolidation of Research Units: AtlantTIC), the Spanish Ministry of Economy and Competitiveness (MINECO), the Fondo Europeo de Desarrollo Regional (FEDER) through Grant No.~TEC2017-88243-R, and the European Union's Horizon 2020 research and innovation programme under the Marie Sk\l{}odowska-Curie grant agreement No 675662 (project QCALL). W. W. gratefully acknowledges support from the National Natural Science Foundation of China (Grants No. 61701539, 61972413, 61901525) and the National Cryptography Development Fund (mmjj20180107, mmjj20180212).

\def\bibsection{\medskip\begin{center}\rule{0.5\columnwidth}{.8pt}\end{center}\medskip} 

\begin{thebibliography}{99}
\bibitem{footnote} We note that a particular USD strategy against the COW protocol has been considered in~\cite{upp_cow2}. In that work, however, the authors do not consider sequential attacks. Also, from the data associated with those signals first measured and then resent by Eve to Bob, he can only estimate two (out of the five) possible visibilities $V_s$.
\bibitem{Vave} The average visibility is defined as $V_{\rm ave}=[p_{\rm click}({\rm D}_{{\rm M}1})-p_{\rm click}({\rm D}_{{\rm M}2})]/[p_{\rm click}({\rm D}_{{\rm M}1})+p_{\rm click}({\rm D}_{{\rm M}2})]$, where the probabilities $p_{\rm click}({\rm D}_{{\rm M}i})=\sum_{s\in{\mathcal S}}P_{s} p_{\rm click}({\rm D}_{{\rm M}i}|s)$ with $i=1,2$. Here, $P_{s}$ denotes the probability that Alice emits the sequence $s$.
\end{thebibliography}
\end{document}